\documentstyle{article}

\begin{document}

\title{THE POINCARE GROUP IS DEDUCED FROM\\
THE LOGIC PROPERTIES OF THE INFORMATION}
\author{Gunn Quznetsov \\
quznets@geocities.com}
\maketitle

\begin{abstract}
The principal time properties - the one-dimensionality and the
irreversibility -, the space metric properties and the spatial-temporal
principles of the theory of the relativity are deduced from three natural
logic properties of the information, obtained by a physics device. Hence,
the transformations of the complete Poincare group are deduced from that.
\end{abstract}

\tableofcontents

\section{LOGIC}

An information, which is obtained from a physics device, can be expressed by
a text.

That is in this article the information is an entity, which can be expressed
by a set of any language sentences.

The sentence $\ll \Theta \gg $ is {\it the true sentence} if and only if $%
\Theta $ \footnote{%
Perhaps, the definition of the truth sentence belongs to A.Tarsky.}.

For example: the sentence $\ll $it rains$\gg $ is the true sentence if and
only if it rains.

The sentence $\ll \Theta \gg $ is {\it the false sentence} if and only if it
is not that $\Theta $.

The sentence $C$ is {\it the conjunction} of the sentences $A$ and $B$ ($%
C=\left( A\&B\right) $), if $C$ is true if and only if $A$ is true and $B$
is true.

The sentence $C$ is {\it the negation} of the sentence $A$ ( $C=\left( \neg
A\right) $), if $C$ is true if and only if $A$ is false.

The function $g\left( x\right) $, which has got the domain on the sentences
set and has got the range of values in the two-elements set $\left\{
0;1\right\} $, is {\it the Bool's function}, if $g\left( x\right) $ follows
to the following conditions:

1. $g\left( \left( A\&B\right) \right) =g\left( A\right) \cdot g\left(
B\right) $;

$2..g\left( \left( \neg A\right) \right) =1-g\left( A\right) $.

Therefore, we shall work into the classical propositional logic approach 
\cite{Mnd}.

For every $A$: $g\left( A\right) \cdot g\left( A\right) =g\left( A\right) $,
since $g\left( x\right) $ has got the range of values in the two-elements
set $\left\{ 0;1\right\} $.

The sentence $A$ is {\it the tautology}, if for every Bool's function $%
g\left( x\right) $: $g\left( A\right) $ $=1$.

For example: because

$g\left( \neg \left( A\&\left( \neg A\right) \right) \right) $ $=1-$ $%
g\left( \left( A\&\left( \neg A\right) \right) \right) =$ $1-g\left(
A\right) \cdot g\left( \left( \neg A\right) \right) $ $=1-$ $g\left(
A\right) \cdot \left( 1-g\left( A\right) \right) =$

$=1-$ $g\left( A\right) $ $+$ $g\left( A\right) \cdot g\left( A\right) $ $=1$%
,

then $\left( \neg \left( A\&\left( \neg A\right) \right) \right) $ is the
tautology.

If $\left( \neg \left( A\&\left( \neg B\right) \right) \right) $ is the
tautology then $B$ is {\it the logic consequence} from $A$.

Let $\widehat{{\bf a}}$ be an object, which may to accept, to retain and/or
to pass any information \footnote{%
The formalization and the self-consistency see in G.Kuznetsov, Physics
Essays, v.4, n.2, (1991), p.157-171.}. The set ${\bf a}$ of the sentences,
which expresses the information of $\widehat{{\bf a}}$, is defined as {\it %
the recorder} of $\widehat{{\bf a}}$. I.e. the expression $\ll $the sentence 
$A$ is an element of ${\bf a}\gg $ denotes: $\ll \widehat{{\bf a}}$ has got
the information, that the event, which can be expressed by the sentence $A$,
happens.$\gg $, or denotes: $\ll \widehat{{\bf a}}$ knows, that $A\gg $. We
write down such expression in abridged type as $\ll {\bf a}^{\bullet }A\gg $.

The following conditions are fulfilled:

I. For every ${\bf a}$ and for every $A$: it is not, that ${\bf a}^{\bullet
}\left( A\&\left( \neg A\right) \right) $, i.e. every recorder does not
contain the contradiction.

II. For every ${\bf a}$, for every $B$ and for every $A$: if $B$ is the
logic consequence from $A$, and ${\bf a}^{\bullet }A$, then ${\bf a}%
^{\bullet }B$.

*III. For every ${\bf a}$, $b$ and for every $A$: if ${\bf a}^{\bullet }\ll 
{\bf b}^{\bullet }A\gg $ then ${\bf a}^{\bullet }A$.

For example: if the device $\widehat{{\bf a}}$ has got the information, that
the device $\widehat{{\bf b}}$ has got the information, that the particle $%
\overleftarrow{\chi }$ mass equals to $7$, then the device $\widehat{{\bf a}}
$ has got the information, that the particle $\overleftarrow{\chi }$ mass
equals to $7$.

\section{TIME}

Let us consider the finite (possible - empty) arrays of the symbols of the
type: ${\bf q}^{\bullet }$.

The array $\alpha $ is {\it the subarray} of the array $\beta $ ($\alpha
\prec \beta $), if $\alpha $ can be obtained from $\beta $ by the deletion
of some (possible - all) elements.

Let us designate: $\left( \alpha \right) ^1$ is $\alpha $, and $\left(
\alpha \right) ^{k+1}$ is $\alpha \left( \alpha \right) ^k$.

Hence, if $k\leq l$, then $\left( \alpha \right) ^k\prec \left( \alpha
\right) ^l$.

The array $\alpha $ is {\it equivalent} to the array $\beta $ ($\alpha \sim
\beta $), if $\alpha $ can be obtained from $\beta $ by the substitution of
the subarray of the type $\left( {\bf a}^{\bullet }\right) ^k$ by the
subarray of the same type ($\left( {\bf a}^{\bullet }\right) ^s$).

In such case:

III. If $\beta \prec \alpha $ or $\beta \sim \alpha $, then for every $K$:

if ${\bf a}^{\bullet }K$, then ${\bf a}^{\bullet }\left( K\&\left( \neg
\left( \left( \alpha A\right) \&\left( \neg \left( \beta A\right) \right)
\right) \right) \right) $.

It is obvious, that III is the refinement of *III.

The number $q$ is {\it the moment}, at which ${\bf a}$ records $B$ by the $%
\kappa -${\it clock} $\left\{ {\bf g}_0,A,{\bf b}_0\right\} $ (the
designation: ${\bf q}$ is $\left[ {\bf a}^{\bullet }B\uparrow {\bf a}%
,\left\{ {\bf g}_0,A,{\bf b}_0\right\} \right] $), if:

1. for every $K$: if ${\bf a}^{\bullet }K$, then ${\bf a}^{\bullet }\left(
K\&\left( \neg \left( \left( {\bf a}^{\bullet }B\right) \&\left( \neg \left( 
{\bf a}^{\bullet }\left( {\bf g}_0^{\bullet }{\bf b}_0^{\bullet }\right) ^q%
{\bf g}_0^{\bullet }A\right) \right) \right) \right) \right) $ and

${\bf a}^{\bullet }\left( K\&\left( \neg \left( \left( {\bf a}^{\bullet
}\left( {\bf g}_0^{\bullet }{\bf b}_0^{\bullet }\right) ^{q+1}{\bf g}%
_0^{\bullet }A\right) \&\left( \neg \left( {\bf a}^{\bullet }B\right)
\right) \right) \right) \right) ;$

2. ${\bf a}^{\bullet }\left( \left( {\bf a}^{\bullet }B\right) \&\left( \neg
\left( {\bf a}^{\bullet }\left( {\bf g}_0^{\bullet }{\bf b}_0^{\bullet
}\right) ^{q+1}{\bf g}_0^{\bullet }A\right) \right) \right) $.

In our world the $\kappa -$clock $\left\{ {\bf g}_0,A,{\bf b}_0\right\} $
accords to the pair ($\widehat{{\bf g}_0}$, $\widehat{{\bf b}_0}$) of the
physics devices, which dispatch the information, expressed by $A$, between
each other. Here the events, expressed by following sentences, happen:

${\bf a}^{\bullet }{\bf g}_0^{\bullet }A$,

${\bf a}^{\bullet }{\bf b}_0^{\bullet }{\bf g}_0^{\bullet }A$,

${\bf a}^{\bullet }{\bf g}_0^{\bullet }{\bf b}_0^{\bullet }{\bf g}%
_0^{\bullet }A$,

${\bf a}^{\bullet }{\bf b}_0^{\bullet }{\bf g}_0^{\bullet }{\bf b}%
_0^{\bullet }{\bf g}_0^{\bullet }A={\bf a}^{\bullet }{\bf b}_0^{\bullet
}\left( {\bf g}_0^{\bullet }{\bf b}_0^{\bullet }\right) ^1{\bf g}_0^{\bullet
}A$,

${\bf a}^{\bullet }\left( {\bf g}_0^{\bullet }{\bf b}_0^{\bullet }\right) ^2%
{\bf g}_0^{\bullet }A$,

${\bf a}^{\bullet }{\bf b}_0^{\bullet }\left( {\bf g}_0^{\bullet }{\bf b}%
_0^{\bullet }\right) ^2{\bf g}_0^{\bullet }A$,

${\bf a}^{\bullet }\left( {\bf g}_0^{\bullet }{\bf b}_0^{\bullet }\right) ^3%
{\bf g}_0^{\bullet }A$, etc.

{\bf Lemma 1 }If

\begin{equation}
q\mbox{ is }\left[ {\bf a}^{\bullet }\alpha B\uparrow {\bf a},\left\{ {\bf g}%
_0,A,{\bf b}_0\right\} \right] ,  \label{b1}
\end{equation}

\begin{equation}
p\mbox{ is }\left[ {\bf a}^{\bullet }\beta B\uparrow {\bf a},\left\{ {\bf g}%
_0,A,{\bf b}_0\right\} \right] ,  \label{b2}
\end{equation}

\begin{equation}
\alpha \prec \beta  \label{b3}
\end{equation}

then

\[
q\leq p 
\]

{\bf Proof:} From (\ref{b2}):

\[
{\bf a}^{\bullet }\left( \left( {\bf a}^{\bullet }\beta B\right) \&\left(
\neg \left( {\bf a}^{\bullet }\left( {\bf g}_0^{\bullet }{\bf b}_0^{\bullet
}\right) ^{\left( p+1\right) }{\bf g_0^{\bullet }}A\right) \right) \right) . 
\]

From above and (\ref{b3}) by III:

\[
{\bf a}^{\bullet }\left( \left( \left( {\bf a}^{\bullet }\beta B\right)
\&\left( \neg \left( {\bf a}^{\bullet }\left( {\bf g}_0^{\bullet }{\bf b}%
_0^{\bullet }\right) ^{\left( p+1\right) }{\bf g_0^{\bullet }}A\right)
\right) \right) \&\left( \neg \left( \left( {\bf a}^{\bullet }\beta B\right)
\&\left( \neg \left( {\bf a}^{\bullet }\alpha B\right) \right) \right)
\right) \right) . 
\]

From above by II:

\[
{\bf a}^{\bullet }\left( \left( {\bf a}^{\bullet }\alpha B\right) \&\left(
\neg \left( {\bf a}^{\bullet }\left( {\bf g}_0^{\bullet }{\bf b}_0^{\bullet
}\right) ^{\left( p+1\right) }{\bf g_0^{\bullet }}A\right) \right) \right) 
\]

From above and (\ref{b1}):

\[
{\bf a}^{\bullet }\left( 
\begin{array}{c}
\left( \left( {\bf a}^{\bullet }\alpha B\right) \&\left( \neg \left( {\bf a}%
^{\bullet }\left( {\bf g}_0^{\bullet }{\bf b}_0^{\bullet }\right) ^{\left(
p+1\right) }{\bf g_0^{\bullet }}A\right) \right) \right) \& \\ 
\left( \neg \left( \left( {\bf a}^{\bullet }\alpha B\right) \&\left( \neg
\left( {\bf a}^{\bullet }\left( {\bf g}_0^{\bullet }{\bf b}_0^{\bullet
}\right) ^q{\bf g_0^{\bullet }}A\right) \right) \right) \right)
\end{array}
\right) . 
\]

From above by II:

\begin{equation}
{\bf a}^{\bullet }\left( \left( \neg \left( {\bf a}^{\bullet }\left( {\bf g}%
_0^{\bullet }{\bf b}_0^{\bullet }\right) ^{\left( p+1\right) }{\bf %
g_0^{\bullet }}A\right) \right) \&\left( {\bf a}^{\bullet }\left( {\bf g}%
_0^{\bullet }{\bf b}_0^{\bullet }\right) ^q{\bf g_0^{\bullet }}A\right)
\right)  \label{b4}
\end{equation}

If $q>p$, that is $q\geq p$, then from (\ref{b4}) by III:

\[
{\bf a}^{\bullet }\left( 
\begin{array}{c}
\left( \left( \neg \left( {\bf a}^{\bullet }\left( {\bf g}_0^{\bullet }{\bf b%
}_0^{\bullet }\right) ^{\left( p+1\right) }{\bf g_0^{\bullet }}A\right)
\right) \&\left( {\bf a}^{\bullet }\left( {\bf g}_0^{\bullet }{\bf b}%
_0^{\bullet }\right) ^q{\bf g_0^{\bullet }}A\right) \right) \& \\ 
\left( \neg \left( \left( {\bf a}^{\bullet }\left( {\bf g}_0^{\bullet }{\bf b%
}_0^{\bullet }\right) ^q{\bf g_0^{\bullet }}A\right) \&\left( \neg \left( 
{\bf a}^{\bullet }\left( {\bf g}_0^{\bullet }{\bf b}_0^{\bullet }\right)
^{\left( p+1\right) }{\bf g_0^{\bullet }}A\right) \right) \right) \right)
\end{array}
\right) . 
\]

From above by II:

\[
{\bf a}^{\bullet }\left( \left( \neg \left( {\bf a}^{\bullet }\left( {\bf g}%
_0^{\bullet }{\bf b}_0^{\bullet }\right) ^{\left( p+1\right) }{\bf %
g_0^{\bullet }}A\right) \right) \&\left( {\bf a}^{\bullet }\left( {\bf g}%
_0^{\bullet }{\bf b}_0^{\bullet }\right) ^{\left( p+1\right) }{\bf %
g_0^{\bullet }}A\right) \right) , 
\]

This is the contradiction of I. Hence, $q\leq p$ $_{{\bf \Box }}$

The following proposition is the Lemma 1 direct consequence:

if $q$ is $\left[ {\bf a}^{\bullet }B\uparrow {\bf a},\left\{ {\bf g}_0,A,%
{\bf b}_0\right\} \right] $ and $p$ is $\left[ {\bf a}^{\bullet }B\uparrow 
{\bf a},\left\{ {\bf g}_0,A,{\bf b}_0\right\} \right] $ then $q=p$. Hence,
the expression $\ll q$ is $\left[ {\bf a}^{\bullet }B\uparrow {\bf a}%
,\left\{ {\bf g}_0,A,{\bf b}_0\right\} \right] \gg $ is equivalent to the
expression $\ll q$ $=$ $\left[ {\bf a}^{\bullet }B\uparrow {\bf a},\left\{ 
{\bf g}_0,A,{\bf b}_0\right\} \right] \gg $.

The $\kappa -$clocks $\left\{ {\bf g}_1,B,{\bf b}_1\right\} $ and $\left\{ 
{\bf g}_2,B,{\bf b}_2\right\} $ have got the {\it identical direction} for $%
{\bf a}$, if the following condition fulfills:

if

\begin{center}
$r${\ $=$ $\left[ {\bf a}^{\bullet }\left( {\bf g}_1^{\bullet }{\bf b}%
_1^{\bullet }\right) ^q{\bf g}_1^{\bullet }B\uparrow {\bf a},\left\{ {\bf g}%
_2,B,{\bf b}_2\right\} \right] $,}\\

$s$ $=$ $\left[ {\bf a}^{\bullet }\left( {\bf g}_1^{\bullet }{\bf b}%
_1^{\bullet }\right) ^p{\bf g}_1^{\bullet }B\uparrow {\bf a},\left\{ {\bf g}%
_2,B,{\bf b}_2\right\} \right] $,\\

$q<p$,\\
\end{center}

then

\begin{center}
$r\leq s$.\\
\end{center}

{\bf Theorem 1. All $\kappa -$clocks have got the identical direction.}

\begin{enumerate}
\item  {\bf Proof:}Let
\end{enumerate}

\[
r=\left[ {\bf a}^{\bullet }\left( {\bf g}_1^{\bullet }{\bf b}_1^{\bullet
}\right) ^q{\bf g}_1^{\bullet }B\uparrow {\bf a},\left\{ {\bf g}_2,B,{\bf b}%
_2\right\} \right] , 
\]

\[
s=\left[ {\bf a}^{\bullet }\left( {\bf g}_1^{\bullet }{\bf b}_1^{\bullet
}\right) ^p{\bf g}_1^{\bullet }B\uparrow {\bf a},\left\{ {\bf g}_2,B,{\bf b}%
_2\right\} \right] , 
\]

\[
q<p. 
\]

In this case:

\[
\left( {\bf g}_1^{\bullet }{\bf b}_1^{\bullet }\right) ^q\prec \left( {\bf g}%
_1^{\bullet }{\bf b}_1^{\bullet }\right) ^p. 
\]

Hence, by Lemma 1:

\[
r\leq s._{{\bf \Box }} 
\]

Therefore, a recorder arranges its own sentences on the moments. And this
order is linear and does not depend from which $\kappa -$clock this order is
settled.

The $\kappa -$clock $\left\{ {\bf g}_2,B,{\bf b}_2\right\} $ is $k$ times%
{\it \ more precise} than the $\kappa -$clock $\left\{ {\bf g}_1,B,{\bf b}%
_1\right\} $ for the recorder ${\bf a}$, if for every $C$ the following
condition fulfills:

if

\begin{center}
$q_1$ $=$ $\left[ {\bf a}^{\bullet }C\uparrow {\bf a},\left\{ {\bf g}_1,B,%
{\bf b}_1\right\} \right] $,\\

$q_2$ $=$ $\left[ {\bf a}^{\bullet }C\uparrow {\bf a},\left\{ {\bf g}_2,B,%
{\bf b}_2\right\} \right] $,\\
\end{center}

then

\begin{center}
$q${$_1$ $<$ }$q${$_2$ $/$ $k$ $<$ }$q${$_1+1$.}\\
\end{center}

The array $\widetilde{H}$ of the $\kappa -$clocks:

\begin{center}
$\left\langle {\left\{ {\bf g}_0,A,{\bf b}_0\right\} ,\ \left\{ {\bf g}_1,A,%
{\bf b}_2\right\} ,...,\left\{ {\bf g}_j,A,{\bf b}_j\right\} ,\ ...\ }%
\right\rangle $\\
\end{center}

is {\it the utter precise $\kappa -$clock} for the recorder ${\bf a}$, if
for every $j$ the natural number $k_j$ exists, for which the $\kappa -$clock 
$\left\{ {\bf g}_j,A,{\bf b}_j\right\} $ is $k_j$ times{\it \ }more precise
than the $\kappa -$clock $\left\{ {\bf g}_{j-1},A,{\bf b}_{j-1}\right\} $.

In this case, if

\[
q_j=\left[ {\bf a}^{\bullet }C\uparrow {\bf a},\left\{ {\bf g}_j,A,{\bf b}%
_j\right\} \right] , 
\]

\[
t=q_0+\sum_{j=1}^\infty \left( q_j-q_{j-1}\cdot k_j\right) /\left( k_1\cdot
k_2\cdot ...\cdot k_j\right) , 
\]

then

\begin{center}
$t$ $=$ $\left[ {\bf a}^{\bullet }C\uparrow {\bf a},\widetilde{H}\right] $.\\
\end{center}

Hence, ''the time'' of the $\kappa -$clocks of type {$\left\{ {\bf g}_j,A,%
{\bf b}_j\right\} $ is the natural number, and ''the time'' of the utter
precise $\kappa -$clocks is the real number.}

\section{SPACE}

Let us denote:

\begin{center}
$\jmath \left( {\bf a}\widetilde{H}\right) \left( {\bf a}^{\bullet }\alpha 
{\bf a}^{\bullet }C\right) =${$\left[ {\bf a}^{\bullet }\alpha {\bf a}%
^{\bullet }C\uparrow {\bf a},\widetilde{H}\right] -\left[ {\bf a}^{\bullet
}C\uparrow {\bf a},\widetilde{H}\right] $.}\\
\end{center}

In our world {$\jmath \left( {\bf a}H\right) \left( {\bf a}^{\bullet }{\bf a}%
_1^{\bullet }{\bf a}_2^{\bullet }{\bf a}^{\bullet }C\right) $ is the time,
at which the information about the event, expressed by sentence }$C$, runs
by the path $\widehat{{\bf a}}$, $\widehat{{\bf a}_1}$, $\widehat{{\bf a}_2}$%
, $\widehat{{\bf a}}$.

Let us denote:

1) for every recorder ${\bf a}$: $\left( {\bf a}\right) ^{\dagger }=\left( 
{\bf a}\right) $;

2) for all recorders series $\alpha $ and $\beta $: $\left( \alpha \beta
\right) ^{\dagger }=\left( \beta \right) ^{\dagger }\left( \alpha \right)
^{\dagger }$.

The set $\Re $ of the recorders is {\it the internally stable system} for
the recorder ${\bf a}$ with the $\kappa -$clock $\widetilde{H}$ (denote: $%
\Re $ is $ISS\left( {\bf a},\widetilde{H}\right) $), if for all sentences $B$
and $C$, for all elements ${\bf a}_1$ and ${\bf a}_2$ of $\Re $ and for all
series $\alpha $, formed by elements of $\Re $, the following conditions are
fulfilled:

1) {$\left[ {\bf a}^{\bullet }{\bf a}_2^{\bullet }{\bf a}_1^{\bullet
}C\uparrow {\bf a},\widetilde{H}\right] -\left[ {\bf a}^{\bullet }{\bf a}%
_1^{\bullet }C\uparrow {\bf a},\widetilde{H}\right] =$}

$=${$\left[ {\bf a}^{\bullet }{\bf a}_2^{\bullet }{\bf a}_1^{\bullet
}B\uparrow {\bf a},\widetilde{H}\right] -\left[ {\bf a}^{\bullet }{\bf a}%
_1^{\bullet }B\uparrow {\bf a},\widetilde{H}\right] $;}

2) {$\jmath \left( {\bf a}\widetilde{H}\right) \left( {\bf a}^{\bullet
}\alpha {\bf a}^{\bullet }C\right) =\jmath \left( {\bf a}\widetilde{H}%
\right) \left( {\bf a}^{\bullet }\alpha ^{\dagger }{\bf a}^{\bullet
}C\right) $.}

In the our world $ISS\left( {\bf a},\widetilde{H}\right) $ accords to the
set of the physics devices, which all are immovable between each other, and
this set does not birl (does not circumvolve).

{\bf Lemma 2.} If $\left\{ {\bf a},{\bf a}_1,{\bf a}_2\right\} $ is $%
ISS\left( {\bf a},\widetilde{H}\right) $, then

\begin{center}
$\left[ {\bf a}^{\bullet }{\bf a}_2^{\bullet }{\bf a}_1^{\bullet }{\bf a}%
_2^{\bullet }C\uparrow {\bf a},\widetilde{H}\right] -\left[ {\bf a}^{\bullet
}{\bf a}_2^{\bullet }C\uparrow {\bf a},\widetilde{H}\right] =$\\

$=\left[ {\bf a}^{\bullet }{\bf a}_1^{\bullet }{\bf a}_2^{\bullet }{\bf a}%
_1^{\bullet }B\uparrow {\bf a},\widetilde{H}\right] -\left[ {\bf a}^{\bullet
}{\bf a}_1^{\bullet }B\uparrow {\bf a},\widetilde{H}\right] $\\
\end{center}

{\bf Proof:}Let us denote:

\begin{center}
$p=\left[ {\bf a}^{\bullet }{\bf a}_1^{\bullet }B\uparrow {\bf a},\widetilde{%
H}\right] ,$

$q=\left[ {\bf a}^{\bullet }{\bf a}_1^{\bullet }{\bf a}_2^{\bullet }{\bf a}%
_1^{\bullet }B\uparrow {\bf a},\widetilde{H}\right] ,$

$r=\left[ {\bf a}^{\bullet }{\bf a}_2^{\bullet }C\uparrow {\bf a},\widetilde{%
H}\right] ,$

$s={\left[ {\bf a}^{\bullet }{\bf a}_2^{\bullet }{\bf a}_1^{\bullet }{\bf a}%
_2^{\bullet }C\uparrow {\bf a},\widetilde{H}\right] .}$

$u=\left[ {\bf a}^{\bullet }{\bf a}_2^{\bullet }{\bf a}_1^{\bullet
}B\uparrow {\bf a},\widetilde{H}\right] ,$

$w={\left[ {\bf a}^{\bullet }{\bf a}_1^{\bullet }{\bf a}_2^{\bullet
}C\uparrow {\bf a},\widetilde{H}\right] }.$
\end{center}

From above by the definition of $ISS$:

\[
u-p=s-w,w-r=q-u. 
\]

From above:

\[
s-r=q-p._{{\bf \Box }} 
\]

Let us denote:

\begin{center}
${\ell }${$\left( {\bf a},\widetilde{H},B\right) \left( {\bf a}_1,{\bf a}%
_2\right) =0.5\cdot \left( \left[ {\bf a}^{\bullet }{\bf a}_1^{\bullet }{\bf %
a}_2^{\bullet }{\bf a}_1^{\bullet }B\uparrow {\bf a},\widetilde{H}\right]
-\left[ {\bf a}^{\bullet }{\bf a}_1^{\bullet }B\uparrow {\bf a},\widetilde{H}%
\right] \right) $.}\\
\end{center}

{\bf Lemma 3. }For all $B$ and $C$: If $\left\{ {\bf a},{\bf a}_1,{\bf a}%
_2\right\} $ is $ISS\left( {\bf a},\widetilde{H}\right) $, then for all $B$
and $C$:

\begin{center}
${\ell }${$\left( {\bf a},\widetilde{H},B\right) \left( {\bf a}_1,{\bf a}%
_2\right) =$}${\ell }${$\left( {\bf a},\widetilde{H},C\right) \left( {\bf a}%
_1,{\bf a}_2\right) $.}\\
\end{center}

{\bf Proof:} Let us denote:

\begin{center}
$p=\left[ {\bf a}^{\bullet }{\bf a}_1^{\bullet }B\uparrow {\bf a},\widetilde{%
H}\right] ,$

$q=\left[ {\bf a}^{\bullet }{\bf a}_1^{\bullet }{\bf a}_2^{\bullet }{\bf a}%
_1^{\bullet }B\uparrow {\bf a},\widetilde{H}\right] ,$

$r=\left[ {\bf a}^{\bullet }{\bf a}_1^{\bullet }C\uparrow {\bf a},\widetilde{%
H}\right] ,$

$s=\left[ {\bf a}^{\bullet }{\bf a}_1^{\bullet }{\bf a}_2^{\bullet }{\bf a}%
_1^{\bullet }C\uparrow {\bf a},\widetilde{H}\right] .$

$u=\left[ {\bf a}^{\bullet }{\bf a}_2^{\bullet }{\bf a}_1^{\bullet
}B\uparrow {\bf a},\widetilde{H}\right] ,$

$w=\left[ {\bf a}^{\bullet }{\bf a}_2^{\bullet }{\bf a}_1^{\bullet
}C\uparrow {\bf a},\widetilde{H}\right] .$
\end{center}

From above by the definition of $ISS$:

\[
u-p=w-r,q-u=s-w. 
\]

From above:

\[
q-p=s-r._{{\bf \Box }} 
\]

Hence, in this case:

\begin{center}
${\ell }${$\left( {\bf a},\widetilde{H},B\right) \left( {\bf a}_1,{\bf a}%
_2\right) =$}${\ell }${$\left( {\bf a},\widetilde{H}\right) \left( {\bf a}_1,%
{\bf a}_2\right) $.}\\
\end{center}

{\bf Theorem 2}: If $\left\{ {\bf a},{\bf a}_1,{\bf a}_2,{\bf a}_3\right\} $
is $ISS\left( {\bf a},\widetilde{H}\right) $, then:

1) ${\ell }${$\left( {\bf a},\widetilde{H}\right) \left( {\bf a}_1,{\bf a}%
_2\right) \geq 0$;}

2) ${\ell }${$\left( {\bf a},\widetilde{H}\right) \left( {\bf a}_1,{\bf a}%
_1\right) =0$;}

3) ${\ell }${$\left( {\bf a},\widetilde{H}\right) \left( {\bf a}_1,{\bf a}%
_2\right) =$}${\ell }${$\left( {\bf a},\widetilde{H}\right) \left( {\bf a}_2,%
{\bf a}_1\right) $;}

4) ${\ell }${$\left( {\bf a},\widetilde{H}\right) \left( {\bf a}_1,{\bf a}%
_2\right) +$}${\ell }${$\left( {\bf a},\widetilde{H}\right) \left( {\bf a}_2,%
{\bf a}_3\right) \geq $}${\ell }${$\left( {\bf a},\widetilde{H}\right)
\left( {\bf a}_1,{\bf a}_3\right) $.}

{\bf Proof: }1) and 2) are the direct consequence from Lemma 1, and 3) -
from Lemma 2.

Let us denote:

\begin{center}
$p=\left[ {\bf a}^{\bullet }{\bf a}_1^{\bullet }C\uparrow {\bf a},\widetilde{%
H}\right] ,$

$q=\left[ {\bf a}^{\bullet }{\bf a}_1^{\bullet }{\bf a}_2^{\bullet }{\bf a}%
_1^{\bullet }C\uparrow {\bf a},\widetilde{H}\right] ,$

$r=\left[ {\bf a}^{\bullet }{\bf a}_1^{\bullet }{\bf a}_3^{\bullet }{\bf a}%
_1^{\bullet }C\uparrow {\bf a},\widetilde{H}\right] ,$

$s=\left[ {\bf a}^{\bullet }{\bf a}_2^{\bullet }{\bf a}_1^{\bullet
}C\uparrow {\bf a},\widetilde{H}\right] .$

$u=\left[ {\bf a}^{\bullet }{\bf a}_2^{\bullet }{\bf a}_3^{\bullet }{\bf a}%
_2^{\bullet }{\bf a}_1^{\bullet }B\uparrow {\bf a},\widetilde{H}\right] ,$

$w=\left[ {\bf a}^{\bullet }{\bf a}_1^{\bullet }{\bf a}_2^{\bullet }{\bf a}%
_3^{\bullet }{\bf a}_2^{\bullet }{\bf a}_1^{\bullet }C\uparrow {\bf a},%
\widetilde{H}\right] .$
\end{center}

From above by the definition of $ISS$:

\[
w-u=q-s, 
\]

Hence,

\[
w-p=\left( q-p\right) +\left( u-s\right) . 
\]

And by Lemma1:

\[
w\geq r. 
\]

Therefore:

\[
\left( q-p\right) +\left( u-s\right) \geq r-p._{{\bf \Box }} 
\]

Thus, {\bf all metric space} \cite{MSP} {\bf axioms are fulfilled for} ${%
\ell }${$\left( {\bf a},\widetilde{H}\right) $ {\bf in the internal stable
system.}}

$B$ happens in {\it the same place} with ${\bf a}_1$ for ${\bf a}$ (denote:$%
\natural \left( {\bf a}\right) \left( {\bf a}_1,B\right) $), if for every
array $\alpha $ and for every sentence $K$ the following condition fulfills:

if ${\bf a}^{\bullet }K$, then ${\bf a}^{\bullet }\left( K\&\left( \neg
\left( \left( \alpha B\right) \&\left( \neg \left( \alpha {\bf a}_1^{\bullet
}B\right) \right) \right) \right) \right) $ .

{\bf Theorem 3} $\natural \left( {\bf a}\right) \left( {\bf a}_1,{\bf a}%
_1^{\bullet }B\right) $.

{\bf Proof: }Since $\alpha {\bf a}_1^{\bullet }\sim \alpha {\bf a}%
_1^{\bullet }{\bf a}_1^{\bullet }$, then by III: if ${\bf a}_1^{\bullet }K$,
then

\[
{\bf a}_1^{\bullet }\left( K\&\left( \neg \left( \left( \alpha {\bf a}%
_1^{\bullet }B\right) \&\left( \neg \left( \alpha {\bf a}_1^{\bullet }{\bf a}%
_1^{\bullet }B\right) \right) \right) \right) \right) ._{{\bf \Box }} 
\]

{\bf Theorem 4.} if

\begin{equation}
\natural \left( {\bf a}\right) \left( {\bf a}_1,B\right) ,  \label{b23}
\end{equation}

\begin{equation}
\natural \left( {\bf a}\right) \left( {\bf a}_2,B\right) ,  \label{b24}
\end{equation}

then

\[
\natural \left( {\bf a}\right) \left( {\bf a}_2,{\bf a}_1^{\bullet }B\right)
. 
\]

{\bf Proof:}Let ${\bf a}^{\bullet }K$. In this case from (\ref{b24}):

\[
{\bf a}^{\bullet }\left( K\&\left( \neg \left( \left( \alpha {\bf a}%
_1^{\bullet }B\right) \&\left( \neg \left( \alpha {\bf a}_1^{\bullet }{\bf a}%
_2^{\bullet }B\right) \right) \right) \right) \right) . 
\]

From (\ref{b23}):

\[
{\bf a}^{\bullet }\left( 
\begin{array}{c}
\left( K\&\left( \neg \left( \left( \alpha {\bf a}_1^{\bullet }B\right)
\&\left( \neg \left( \alpha {\bf a}_1^{\bullet }{\bf a}_2^{\bullet }B\right)
\right) \right) \right) \right) \& \\ 
\left( \neg \left( \left( \alpha {\bf a}_1^{\bullet }{\bf a}_2^{\bullet
}B\right) \&\left( \neg \left( \alpha {\bf a}_1^{\bullet }{\bf a}_2^{\bullet
}{\bf a}_1^{\bullet }B\right) \right) \right) \right)
\end{array}
\right) \mbox{.} 
\]

From above by II:

\[
{\bf a}^{\bullet }\left( K\&\left( \neg \left( \left( \alpha {\bf a}%
_1^{\bullet }B\right) \&\left( \neg \left( \alpha {\bf a}_1^{\bullet }{\bf a}%
_2^{\bullet }{\bf a}_1^{\bullet }B\right) \right) \right) \right) \right) . 
\]

From above by III:

\[
{\bf a}^{\bullet }\left( 
\begin{array}{c}
\left( K\&\left( \neg \left( \left( \alpha {\bf a}_1^{\bullet }B\right)
\&\left( \neg \left( \alpha {\bf a}_1^{\bullet }{\bf a}_2^{\bullet }{\bf a}%
_1^{\bullet }B\right) \right) \right) \right) \right) \& \\ 
\left( \neg \left( \left( \alpha {\bf a}_1^{\bullet }{\bf a}_2^{\bullet }%
{\bf a}_1^{\bullet }B\right) \&\left( \neg \left( \alpha {\bf a}_2^{\bullet }%
{\bf a}_1^{\bullet }B\right) \right) \right) \right)
\end{array}
\right) \mbox{.} 
\]

From above by II:

\[
{\bf a}^{\bullet }\left( K\&\left( \neg \left( \left( \alpha {\bf a}%
_1^{\bullet }B\right) \&\left( \neg \left( \alpha {\bf a}_2^{\bullet }{\bf a}%
_1^{\bullet }B\right) \right) \right) \right) \right) ._{{\bf \Box }} 
\]

{\bf Lemma 4.} If

\begin{equation}
\natural \left( {\bf a}\right) \left( {\bf a}_1,B\right) ,  \label{b25}
\end{equation}

\begin{equation}
t\ =\ \left[ {\bf a}^{\bullet }\alpha B\uparrow {\bf a},\widetilde{H}\right]
,  \label{b26}
\end{equation}

{then }

\[
t\ =\ \left[ {\bf a}^{\bullet }\alpha {\bf a}_1^{\bullet }B\uparrow {\bf a},%
\widetilde{H}\right] \mbox{.} 
\]

{\bf Proof:} Let us denote:

\[
t_j=\left[ {\bf a}^{\bullet }\alpha B\uparrow {\bf a},\left\{ {\bf g}_j,A,%
{\bf b}_j\right\} \right] . 
\]

Hence:

\[
{\bf a}^{\bullet }\left( \left( {\bf a}^{\bullet }\alpha B\right) \&\left(
\neg \left( {\bf a}^{\bullet }\left( {\bf g}_j^{\bullet }{\bf b}_j^{\bullet
}\right) ^{t_j+1}{\bf g}_j^{\bullet }A\right) \right) \right) , 
\]

and from (\ref{b25}):

\[
{\bf a}^{\bullet }\left( 
\begin{array}{c}
\left( \left( {\bf a}^{\bullet }\alpha B\right) \&\left( \neg \left( {\bf a}%
^{\bullet }\left( {\bf g}_j^{\bullet }{\bf b}_j^{\bullet }\right) ^{t_j+1}%
{\bf g}_j^{\bullet }A\right) \right) \right) \& \\ 
\left( \neg \left( \left( {\bf a}^{\bullet }\alpha B\right) \&\left( \neg
\left( {\bf a}^{\bullet }\alpha {\bf a}_1^{\bullet }B\right) \right) \right)
\right)
\end{array}
\right) \mbox{.} 
\]

From above by II:

\begin{equation}
{\bf a}^{\bullet }\left( \left( {\bf a}^{\bullet }\alpha {\bf a}_1^{\bullet
}B\right) \&\left( \neg \left( {\bf a}^{\bullet }\left( {\bf g}_j^{\bullet }%
{\bf b}_j^{\bullet }\right) ^{t_j+1}{\bf g}_j^{\bullet }A\right) \right)
\right) ,  \label{b27}
\end{equation}

Let ${\bf a}^{\bullet }K$. In this case from (\ref{b26}):

\[
{\bf a}^{\bullet }\left( K\&\left( \neg \left( \left( {\bf a}^{\bullet
}\alpha B\right) \&\left( \neg \left( {\bf a}^{\bullet }\left( {\bf g}%
_j^{\bullet }{\bf b}_j^{\bullet }\right) ^{t_j}{\bf g}_j^{\bullet }A\right)
\right) \right) \right) \right) . 
\]

Hence, by III:

\[
{\bf a}^{\bullet }\left( 
\begin{array}{c}
\left( K\&\left( \neg \left( \left( {\bf a}^{\bullet }\alpha B\right)
\&\left( \neg \left( {\bf a}^{\bullet }\left( {\bf g}_j^{\bullet }{\bf b}%
_j^{\bullet }\right) ^{t_j}{\bf g}_j^{\bullet }A\right) \right) \right)
\right) \right) \& \\ 
\left( \neg \left( \left( {\bf a}^{\bullet }\alpha {\bf a}_1^{\bullet
}B\right) \&\left( \neg \left( {\bf a}^{\bullet }\alpha B\right) \right)
\right) \right)
\end{array}
\right) \mbox{.} 
\]

From above by II:

\begin{equation}
{\bf a}^{\bullet }\left( K\&\left( \neg \left( \left( {\bf a}^{\bullet
}\alpha {\bf a}_1^{\bullet }B\right) \&\left( \neg \left( {\bf a}^{\bullet
}\left( {\bf g}_j^{\bullet }{\bf b}_j^{\bullet }\right) ^{t_j}{\bf g}%
_j^{\bullet }A\right) \right) \right) \right) \right) .  \label{b28}
\end{equation}

From above and from (\ref{b25}):

\[
{\bf a}^{\bullet }\left( 
\begin{array}{c}
\left( K\&\left( \neg \left( \left( {\bf a}^{\bullet }\left( {\bf g}%
_j^{\bullet }{\bf b}_j^{\bullet }\right) ^{t_j+1}{\bf g}_j^{\bullet
}A\right) \&\left( \neg \left( {\bf a}^{\bullet }\alpha B\right) \right)
\right) \right) \right) \& \\ 
\left( \neg \left( \left( {\bf a}^{\bullet }\alpha B\right) \&\left( \neg
\left( {\bf a}^{\bullet }\alpha {\bf a}_1^{\bullet }B\right) \right) \right)
\right)
\end{array}
\right) \mbox{.} 
\]

From above by II:

\[
{\bf a}^{\bullet }\left( K\&\left( \neg \left( \left( {\bf a}^{\bullet
}\left( {\bf g}_j^{\bullet }{\bf b}_j^{\bullet }\right) ^{t_j+1}{\bf g}%
_j^{\bullet }A\right) \&\left( \neg \left( {\bf a}^{\bullet }\alpha {\bf a}%
_1^{\bullet }B\right) \right) \right) \right) \right) . 
\]

From above and from (\ref{b27}), (\ref{b28}) for all $j$:

\[
t_j=\left[ {\bf a}^{\bullet }\alpha {\bf a}_1^{\bullet }B\uparrow {\bf a}%
,\left\{ {\bf g}_j,A,{\bf b}_j\right\} \right] . 
\]

Therefore,

\[
t\ =\ \left[ {\bf a}^{\bullet }\alpha {\bf a}_1^{\bullet }B\uparrow {\bf a},%
\widetilde{H}\right] \mbox{.}_{{\bf \Box }} 
\]

{\bf Theorem 5.} If $\left\{ {\bf a},{\bf a}_1,{\bf a}_2\right\} $ is $%
ISS\left( {\bf a},\widetilde{H}\right) $,

\begin{equation}
\natural \left( {\bf a}\right) \left( {\bf a}_1,B\right) ,  \label{b29}
\end{equation}

\begin{equation}
\natural \left( {\bf a}\right) \left( {\bf a}_2,B\right) ,  \label{b30}
\end{equation}

then

\[
\ell \left( {\bf a},\widetilde{H}\right) \left( {\bf a}_1,{\bf a}_2\right) =0%
\mbox{.} 
\]

{\bf Proof:} Let us denote:

\[
t\ =\ \left[ {\bf a}^{\bullet }B\uparrow {\bf a},\widetilde{H}\right] . 
\]

From above by Lemma 4:

from (\ref{b29}):

\[
t\ =\ \left[ {\bf a}^{\bullet }{\bf a}_1^{\bullet }B\uparrow {\bf a},%
\widetilde{H}\right] \mbox{,} 
\]

from (\ref{b30}):

\[
t=\left[ {\bf a}^{\bullet }{\bf a}_1^{\bullet }{\bf a}_2^{\bullet }B\uparrow 
{\bf a},\widetilde{H}\right] , 
\]

again from (\ref{b29}):

\[
t=\left[ {\bf a}^{\bullet }{\bf a}_1^{\bullet }{\bf a}_2^{\bullet }{\bf a}%
_1^{\bullet }B\uparrow {\bf a},\widetilde{H}\right] . 
\]

Therefore:

\[
\ell \left( {\bf a},\widetilde{H}\right) \left( {\bf a}_1,{\bf a}_2\right)
=0.5\cdot \left( t-t\right) =0\mbox{.}_{{\bf \Box }} 
\]

{\bf Theorem 6.} if $\left\{ {\bf a}_1,{\bf a}_2,{\bf a}_3\right\} $ is $%
ISS\left( {\bf a},\widetilde{H}\right) $ and the sentence $B$ exists, for
which:

\begin{equation}
\natural \left( {\bf a}\right) \left( {\bf a}_1,B\right) ,  \label{b31}
\end{equation}

\begin{equation}
\natural \left( {\bf a}\right) \left( {\bf a}_2,B\right) ,  \label{b32}
\end{equation}

then

\[
{\ell \left( {\bf a},\widetilde{H}\right) \left( {\bf a}_3,{\bf a}_2\right)
=\ell \left( {\bf a},\widetilde{H}\right) \left( {\bf a}_3,{\bf a}_1\right) .%
} 
\]

{\bf Proof:} By Theorem 5 from (\ref{b31}) and (\ref{b32}):

\begin{equation}
\ell \left( {\bf a},\widetilde{H}\right) \left( {\bf a}_1,{\bf a}_2\right)
=0;  \label{b33}
\end{equation}

By Theorem 2:

\[
{\ell \left( {\bf a},\widetilde{H}\right) \left( {\bf a}_1,{\bf a}_2\right)
+\ell \left( {\bf a},\widetilde{H}\right) \left( {\bf a}_2,{\bf a}_3\right)
\geq \ell \left( {\bf a},\widetilde{H}\right) \left( {\bf a}_1,{\bf a}%
_3\right) }\mbox{,} 
\]

hence, from (\ref{b33}):

\[
{\ell \left( {\bf a},\widetilde{H}\right) \left( {\bf a}_2,{\bf a}_3\right)
\geq \ell \left( {\bf a},\widetilde{H}\right) \left( {\bf a}_1,{\bf a}%
_3\right) }\mbox{,} 
\]

hence, by Theorem 2:

\begin{equation}
{\ell \left( {\bf a},\widetilde{H}\right) \left( {\bf a}_3,{\bf a}_2\right)
\geq \ell \left( {\bf a},\widetilde{H}\right) \left( {\bf a}_1,{\bf a}%
_3\right) }\mbox{.}  \label{b34}
\end{equation}

From

\[
{\ell \left( {\bf a},\widetilde{H}\right) \left( {\bf a}_3,{\bf a}_1\right)
+\ell \left( {\bf a},\widetilde{H}\right) \left( {\bf a}_1,{\bf a}_2\right)
\geq \ell \left( {\bf a},\widetilde{H}\right) \left( {\bf a}_3,{\bf a}%
_2\right) }\mbox{:} 
\]

\[
{\ell \left( {\bf a},\widetilde{H}\right) \left( {\bf a}_3,{\bf a}_1\right)
\geq \ell \left( {\bf a},\widetilde{H}\right) \left( {\bf a}_3,{\bf a}%
_2\right) }\mbox{.} 
\]

From above and from (\ref{b34}):

\[
{\ell \left( {\bf a},\widetilde{H}\right) \left( {\bf a}_3,{\bf a}_1\right)
=\ell \left( {\bf a},\widetilde{H}\right) \left( {\bf a}_3,{\bf a}_2\right) }%
\mbox{.}_{{\bf \Box }} 
\]

The real number $t$ is the moment of $B$ for {\it the frame of reference} $%
\left( \Re {\bf a}\widetilde{H}\right) $

(denote: $t=\left[ B\mid \Re {\bf a}\widetilde{H}\right] $), if

1) $\Re $ is $ISS\left( {\bf a},\widetilde{H}\right) $,

2) the recorder ${\bf b}$ exists, for which: ${\bf b}\in \Re $ and $\natural
\left( {\bf a}\right) \left( {\bf b},B\right) $),

3) $t=${$\left[ {\bf a}^{\bullet }B\uparrow {\bf a},\widetilde{H}\right] -$}$%
{\ell }${$\left( {\bf a},\widetilde{H}\right) \left( {\bf a},{\bf b}\right) $%
.}

{\bf Lemma 5.}

\[
{\left[ {\bf a}^{\bullet }B\uparrow {\bf a},\widetilde{H}\right] =}\left[ 
{\bf a}^{\bullet }B\mid \Re {\bf a}\widetilde{H}\right] \mbox{.} 
\]

{\bf Proof:} Let $\Re $ be $ISS\left( {\bf a},\widetilde{H}\right) $, ${\bf a%
}_1\in \Re $ and

\begin{equation}
\natural \left( {\bf a}\right) \left( {\bf a}_1,{\bf a}^{\bullet }B\right) %
\mbox{.}  \label{b35}
\end{equation}

By Theorem 3:

\[
\natural \left( {\bf a}\right) \left( {\bf a},{\bf a}^{\bullet }B\right) %
\mbox{.} 
\]

From above and from (\ref{b35}) by Theorem 5:

\[
\ell \left( {\bf a},\widetilde{H}\right) \left( {\bf a},{\bf a}_1\right) =0%
\mbox{,} 
\]

hence,

\[
\left[ {\bf a}^{\bullet }B\mid \Re {\bf a}\widetilde{H}\right] ={\left[ {\bf %
a}^{\bullet }B\uparrow {\bf a},\widetilde{H}\right] -}\ell \left( {\bf a},%
\widetilde{H}\right) \left( {\bf a},{\bf a}_1\right) ={\left[ {\bf a}%
^{\bullet }B\uparrow {\bf a},\widetilde{H}\right] }\mbox{.}_{{\bf \Box }} 
\]

The real number $z$ is t{\it he distance} between $B$ and $C$ for the frame
reference $\left( \Re {\bf a}\widetilde{H}\right) $

(denote: $z={\ell }\left( \Re {\bf a}\widetilde{H}\right) \left( B,C\right) $%
, if

1) $\Re $ is $ISS\left( {\bf a},\widetilde{H}\right) $,

2) the recorders ${\bf a}_1$ and ${\bf a}_2$ exist, for which: ${\bf a}_1\in
\Re $, ${\bf a}_2\in \Re $, $\natural \left( {\bf a}\right) \left( {\bf a}%
_1,B\right) $) and $\natural \left( {\bf a}\right) \left( {\bf a}_2,C\right) 
$),

3) $z={\ell }${$\left( {\bf a},\widetilde{H}\right) \left( {\bf a}_2,{\bf a}%
_1\right) $.}

From Theorem 2: {\bf all axioms of the metric space fulfill for such
distance.}

\section{RELATIVITY}

The recorders ${\bf a}_1$ and ${\bf a}_2$ {\it receive the information,
expressed by} $B$, {\it identically} for the recorder ${\bf a}$, if $\ll
\natural \left( {\bf a}\right) \left( {\bf a}_2,{\bf a}_1^{\bullet }B\right)
\gg $ is consequence of $\ll \natural \left( {\bf a}\right) \left( {\bf a}_1,%
{\bf a}_2^{\bullet }B\right) \gg $ and vice versa.

The recorders set is {\it the homogeneous space}, if all elements of this
set receive every information identically.

The real number $c$ is {\it the propagation velocity of the information},
expressed by $B$, to the recorder ${\bf a}_1$ for the frame reference $%
\left( \Re {\bf a}\widetilde{H}\right) $, if

\begin{center}
$c=\left( {\ell }\left( \Re {\bf a}\widetilde{H}\right) \left( B,{\bf a}%
_1^{\bullet }B\right) \right) /\left( \left[ {\bf a}_1^{\bullet }B\mid \Re 
{\bf a}\widetilde{H}\right] -\left[ B\mid \Re {\bf a}\widetilde{H}\right]
\right) $.\\
\end{center}

{\bf Theorem 7.} In all homogenous spaces:

\begin{center}
$c=1$.\\
\end{center}

{\bf Proof:} Let $c$ be the propagation velocity of the information,
expressed by $B$, to the recorder ${\bf a}_1$ for the frame reference $%
\left( \Re {\bf a}\widetilde{H}\right) $. That is: if

\[
\Re \mbox{ is }ISS\left( {\bf a},\widetilde{H}\right) , 
\]

\begin{equation}
z={\ell }\left( \Re {\bf a}\widetilde{H}\right) \left( B,{\bf a}_1^{\bullet
}B\right) \mbox{,}  \label{b36}
\end{equation}

\begin{equation}
t_1=\left[ B\mid \Re {\bf a}\widetilde{H}\right] \mbox{,}  \label{b37}
\end{equation}

\begin{equation}
t_2=\left[ {\bf a}_1^{\bullet }B\mid \Re {\bf a}\widetilde{H}\right] ,
\label{b38}
\end{equation}

rhen

\begin{equation}
c=z/\left( t_2-t_1\right) \mbox{.}  \label{b39}
\end{equation}

From (\ref{b36}): the elements ${\bf b}_1$ and ${\bf b}_2$ of $\Re $ exist,
for which:

\begin{equation}
\natural \left( {\bf a}\right) \left( {\bf b}_1,B\right) \mbox{,}
\label{b40}
\end{equation}

\begin{equation}
\natural \left( {\bf a}\right) \left( {\bf b}_2,{\bf a}_2^{\bullet }B\right) %
\mbox{,}  \label{b41}
\end{equation}

\begin{equation}
z={\ell \left( {\bf a},\widetilde{H}\right) \left( {\bf b}_1,{\bf b}%
_2\right) }\mbox{.}  \label{b42}
\end{equation}

From (\ref{b37}) and (\ref{b38}): the elements ${\bf b}_1^{\prime }$ and $%
{\bf b}_2^{\prime }$ of $\Re $ exist, for which:

\begin{equation}
\natural \left( {\bf a}\right) \left( {\bf b}_1^{\prime },B\right) \mbox{,}
\label{b43}
\end{equation}

\begin{equation}
\natural \left( {\bf a}\right) \left( {\bf b}_2^{\prime },{\bf a}_2^{\bullet
}B\right) ,  \label{b44}
\end{equation}

\begin{equation}
t_1={\left[ {\bf a}^{\bullet }B\uparrow {\bf a},\widetilde{H}\right] -\ell
\left( {\bf a},\widetilde{H}\right) \left( {\bf a},{\bf b}_1^{\prime
}\right) }\mbox{,}  \label{b45}
\end{equation}

\begin{equation}
t_2={\left[ {\bf a}^{\bullet }{\bf a}_2^{\bullet }B\uparrow {\bf a},%
\widetilde{H}\right] -\ell \left( {\bf a},\widetilde{H}\right) \left( {\bf a}%
,{\bf b}_2^{\prime }\right) }\mbox{.}  \label{b46}
\end{equation}

From (\ref{b36}), (\ref{b40}), (\ref{b43}) by Theorem 6:

\begin{equation}
{\ell \left( {\bf a},\widetilde{H}\right) \left( {\bf a},{\bf b}_1\right)
=\ell \left( {\bf a},\widetilde{H}\right) \left( {\bf a},{\bf b}_1^{\prime
}\right) }\mbox{.}  \label{b47}
\end{equation}

Like to above from (\ref{b36}), (\ref{b41}), (\ref{b44}):

\begin{equation}
{\ell \left( {\bf a},\widetilde{H}\right) \left( {\bf a},{\bf b}_2\right)
=\ell \left( {\bf a},\widetilde{H}\right) \left( {\bf a},{\bf b}_2^{\prime
}\right) }\mbox{.}  \label{b48}
\end{equation}

From (\ref{b45}), (\ref{b40}), (\ref{b47}) by Lemma 4:

\begin{equation}
t_1={\left[ {\bf a}^{\bullet }{\bf b}_1^{\bullet }B\uparrow {\bf a},%
\widetilde{H}\right] -\ell \left( {\bf a},\widetilde{H}\right) \left( {\bf a}%
,{\bf b}_1\right) }\mbox{.}  \label{b49}
\end{equation}

From (\ref{b41}) by Lemma 4:

\begin{equation}
{\left[ {\bf a}^{\bullet }{\bf a}_2^{\bullet }B\uparrow {\bf a},\widetilde{H}%
\right] =\left[ {\bf a}^{\bullet }{\bf b}_2^{\bullet }{\bf a}_2^{\bullet
}B\uparrow {\bf a},\widetilde{H}\right] }\mbox{.}  \label{b50}
\end{equation}

By Lemma 1:

\begin{equation}
{\left[ {\bf a}^{\bullet }{\bf b}_2^{\bullet }{\bf a}_2^{\bullet }B\uparrow 
{\bf a},\widetilde{H}\right] \geq \left[ {\bf a}^{\bullet }{\bf b}%
_2^{\bullet }B\uparrow {\bf a},\widetilde{H}\right] }\mbox{.}  \label{b51}
\end{equation}

From (\ref{b41}):

\[
\natural \left( {\bf a}\right) \left( {\bf a}_2,{\bf b}_2^{\bullet }B\right) %
\mbox{.} 
\]

From above by Lemma 4:

\begin{equation}
{\left[ {\bf a}^{\bullet }{\bf a}_2^{\bullet }{\bf b}_2^{\bullet }B\uparrow 
{\bf a},\widetilde{H}\right] =\left[ {\bf a}^{\bullet }{\bf b}_2^{\bullet
}B\uparrow {\bf a},\widetilde{H}\right] }\mbox{.}  \label{b52}
\end{equation}

Again by Lemma 1:

\[
{\left[ {\bf a}^{\bullet }{\bf a}_2^{\bullet }{\bf b}_2^{\bullet }B\uparrow 
{\bf a},\widetilde{H}\right] \geq \left[ {\bf a}^{\bullet }{\bf a}%
_2^{\bullet }B\uparrow {\bf a},\widetilde{H}\right] }\mbox{.} 
\]

From above and from (\ref{b52}), (\ref{b50}), (\ref{b51}):

\[
{\left[ {\bf a}^{\bullet }{\bf a}_2^{\bullet }B\uparrow {\bf a},\widetilde{H}%
\right] \geq \left[ {\bf a}^{\bullet }{\bf b}_2^{\bullet }B\uparrow {\bf a},%
\widetilde{H}\right] \geq \left[ {\bf a}^{\bullet }{\bf a}_2^{\bullet
}B\uparrow {\bf a},\widetilde{H}\right] }\mbox{,} 
\]

hence,

\[
{\left[ {\bf a}^{\bullet }{\bf a}_2^{\bullet }B\uparrow {\bf a},\widetilde{H}%
\right] =\left[ {\bf a}^{\bullet }{\bf b}_2^{\bullet }B\uparrow {\bf a},%
\widetilde{H}\right] }\mbox{.} 
\]

From above and from (\ref{b46}), (\ref{b48}):

\[
t_2={\left[ {\bf a}^{\bullet }{\bf b}_2^{\bullet }B\uparrow {\bf a},%
\widetilde{H}\right] -\ell \left( {\bf a},\widetilde{H}\right) \left( {\bf a}%
,{\bf b}_2\right) }\mbox{.} 
\]

From above and from (\ref{b40}) by Lemma 4:

\begin{equation}
t_2={\left[ {\bf a}^{\bullet }{\bf b}_2^{\bullet }{\bf b}_1^{\bullet
}B\uparrow {\bf a},\widetilde{H}\right] -\ell \left( {\bf a},\widetilde{H}%
\right) \left( {\bf a},{\bf b}_2\right) }\mbox{.}  \label{b53}
\end{equation}

Let us denote:

\begin{equation}
u={\left[ {\bf a}^{\bullet }C\uparrow {\bf a},\widetilde{H}\right] }\mbox{,}
\label{b54}
\end{equation}

\begin{equation}
d={\left[ {\bf a}^{\bullet }{\bf b}_1^{\bullet }{\bf a}^{\bullet }C\uparrow 
{\bf a},\widetilde{H}\right] },  \label{b55}
\end{equation}

\begin{equation}
w={\left[ {\bf a}^{\bullet }{\bf b}_2^{\bullet }{\bf a}^{\bullet }C\uparrow 
{\bf a},\widetilde{H}\right] },  \label{b56}
\end{equation}

\begin{equation}
j={\left[ {\bf a}^{\bullet }{\bf b}_2^{\bullet }{\bf b}_1^{\bullet }{\bf a}%
^{\bullet }C\uparrow {\bf a},\widetilde{H}\right] },  \label{b57}
\end{equation}

\[
q={\left[ {\bf a}^{\bullet }{\bf b}_1^{\bullet }{\bf b}_2^{\bullet }{\bf a}%
^{\bullet }C\uparrow {\bf a},\widetilde{H}\right] }, 
\]

\begin{equation}
p={\left[ {\bf a}^{\bullet }{\bf b}_1^{\bullet }{\bf b}_2^{\bullet }{\bf b}%
_1^{\bullet }{\bf a}^{\bullet }C\uparrow {\bf a},\widetilde{H}\right] },
\label{b58}
\end{equation}

\[
r={\left[ {\bf a}^{\bullet }{\bf b}_2^{\bullet }{\bf b}_1^{\bullet }{\bf b}%
_2^{\bullet }{\bf a}^{\bullet }C\uparrow {\bf a},\widetilde{H}\right] }%
\mbox{.} 
\]

Since $\Re $ is $ISS\left( {\bf a},\widetilde{H}\right) $, then

\begin{equation}
q-w=p-j\mbox{,}  \label{b59}
\end{equation}

\begin{equation}
j=q\mbox{.}  \label{b60}
\end{equation}

And from (\ref{b53}), (\ref{b49}), (\ref{b55}), (\ref{b57}):

\[
\left( t_2+{\ell \left( {\bf a},\widetilde{H}\right) \left( {\bf a},{\bf b}%
_2\right) }\right) -\left( t_1+{\ell \left( {\bf a},\widetilde{H}\right)
\left( {\bf a},{\bf b}_1\right) }\right) =j-d\mbox{,} 
\]

hence,

\begin{equation}
t_2-t_1=j-d-{\ell \left( {\bf a},\widetilde{H}\right) \left( {\bf a},{\bf b}%
_2\right) +\ell \left( {\bf a},\widetilde{H}\right) \left( {\bf a},{\bf b}%
_1\right) }\mbox{.}  \label{b61}
\end{equation}

From (\ref{b54}), (\ref{b55}), (\ref{b56}) by Lemma 1:

\[
{\ell \left( {\bf a},\widetilde{H}\right) \left( {\bf a},{\bf b}_2\right)
=0.5\cdot }\left( w-u\right) \mbox{, }{\ell \left( {\bf a},\widetilde{H}%
\right) \left( {\bf a},{\bf b}_1\right) =0.5\cdot }\left( d-u\right) \mbox{.}
\]

From above and from (\ref{b59}), (\ref{b60}), (\ref{b61}):

\[
t_2-t_1=0.5\cdot \left( \left( j-d\right) +\left( j-w\right) \right)
=0.5\cdot \left( j-d+p-j\right) =0.5\cdot \left( p-d\right) \mbox{.} 
\]

From (\ref{b58}), (\ref{b55}), (\ref{b42}):

\[
z=0.5\cdot \left( p-d\right) \mbox{.} 
\]

Therefore,

\[
z=t_2-t_1\mbox{.}_{{\bf \Box }} 
\]

That is{\bf \ in every homogenous space the propagation velocity of every
information to every recorder for every frame reference equals to }$1$.

From this theorem: in all homogenous spaces: ({\bf the time irreversibility})

\begin{center}
$\left[ {\bf a}_1^{\bullet }B\mid \Re {\bf a}\widetilde{H}\right] \geq
\left[ B\mid \Re {\bf a}\widetilde{H}\right] $.\\
\end{center}

Therefore, in every homogenous space: {\bf nobody can learn about that, what
any event occurred, before, than it occurred.}

By the Urysohn theorem \cite{MSP1}: every homogenous space is homeonorph to
some set into the real Hilbert space. If this homomorphism is not the
identity transformation, then $\Re $ is the noneuclidean space. In this case
some variant of General Relativity Theory can be constructed into this
''space-time''. If this homomorphism is some identity transformation, then $%
\Re $ is the Euclidean space. In this case some {\it coordinates system }$%
R^\mu $ exists, for which the following condition fulfills:

for all elements ${\bf a}_1$ and ${\bf a}_2$ of $\Re $ the points $%
\overrightarrow{x_1}$ and $\overrightarrow{x_2}$ of $R^\mu $ exist, for
which:

\begin{center}
${\ell }${$\left( {\bf a},\widetilde{H}\right) \left( {\bf a}_k,{\bf a}%
_s\right) =\left( \sum_{j=1}^\mu \left( x_{s,j}-x_{k,j}\right) ^2\right)
^{0.5}$.}\\
\end{center}

In this case $R^\mu $ is denoted as the coordinates system of the frame
reference $\left( \Re {\bf a}\widetilde{H}\right) $, and the numbers $%
\left\langle x_{k,1},x_{k,2},\ldots ,x_{k,\mu }\right\rangle $ - as {\it the
coordinates of the recorder} ${\bf a}_k$ in $R^\mu $.

{\bf The coordinate system of the frame reference can be determined accurate
to the transformations of the replacement, the rotating and the inversion.}

$B$ has got {\it the coordinates} $\left\langle x_1,x_2,\ldots ,x_\mu
\right\rangle $ in the coordinate system $R^\mu $ of the frame reference $%
\left( \Re {\bf a}\widetilde{H}\right) $, if the recorder ${\bf b}$ exists,
for which: ${\bf b}\in \Re $, $\natural \left( {\bf a}\right) \left( {\bf b}%
,B\right) $ and the coordinates of ${\bf b}$ in $R^\mu $ are $\left\langle
x_1,x_2,\ldots ,x_\mu \right\rangle $.

The recorder ${\bf b}$ has got {\it the coordinates} $\left\langle
x_1,x_2,\ldots ,x_\mu \right\rangle $ in the coordinate system $R^\mu $ in
the moment $t$ of the frame reference $\left( \Re {\bf a}\widetilde{H}%
\right) $, if for every $B$ the following condition fulfills:

if $t=\left[ {\bf b}^{\bullet }B\mid \Re {\bf a}\widetilde{H}\right] $, then 
$\ll {\bf b}^{\bullet }B\gg $ has got the coordinates $\left\langle
x_1,x_2,\ldots ,x_\mu \right\rangle $ in the coordinate system $R^\mu $ of
the frame reference $\left( \Re {\bf a}\widetilde{H}\right) $.

From Theorem 9: For all real numbers $v$ ($\left| v\right| <1$) and $l$, for
the coordinate system $R^\mu $ of the frame reference $\left( \Re {\bf a}%
\widetilde{H}\right) $, if at every moment $t$ the coordinates of:

${\bf b}$ are: $\left\langle x_{{\bf b},1}+v\cdot t,x_{{\bf b},2},x_{{\bf b}%
,3},\ldots ,x_{{\bf b},\mu }\right\rangle $;

${\bf a}_0$ are: $\left\langle x_{0,1}+v\cdot t,x_{0,2},x_{0,3},\ldots
,x_{0,\mu }\right\rangle $;

${\bf b}_0$ are: $\left\langle x_{0,1}+v\cdot t,x_{0,2}+l,x_{0,3},\ldots
,x_{0,\mu }\right\rangle $; and

$t_C=\left[ {\bf b}^{\bullet }C\mid \Re {\bf a}\widetilde{H}\right] $;

$t_D=\left[ {\bf b}^{\bullet }D\mid \Re {\bf a}\widetilde{H}\right] $;

$q_C$ $=$ $\left[ {\bf b}^{\bullet }C\uparrow {\bf b},\left\{ {\bf g}_0,A,%
{\bf b}_0\right\} \right] $;

$q_D$ $=$ $\left[ {\bf b}^{\bullet }D\uparrow {\bf b},\left\{ {\bf g}_0,A,%
{\bf b}_0\right\} \right] $,

then

\begin{center}
$\lim_{l\rightarrow 0}2\cdot l/\sqrt{\left( 1-v^2\right) }\cdot \left(
\left( q_D-q_C\right) /\left( t_D-t_C\right) \right) =1$.\\
\end{center}

If denote: $q_D^{st}=$ $q_D$ and $q_C^{st}=q_C$ for $v=0$, then

\begin{center}
$\lim_{l\rightarrow 0}2\cdot l\cdot \left( \left( q_D^{st}-q_C^{st}\right)
/\left( t_D-t_C\right) \right) =1$.\\
\end{center}

Therefore:

\begin{center}
$q_D-q_C=\left( q_D^{st}-q_C^{st}\right) \cdot \sqrt{\left( 1-v^2\right) }$.%
\\
\end{center}

Hence,{\bf \ a }$\kappa -${\bf clock, which moves with velocity }$v${\bf ,
runs }$\left( 1-v^2\right) ^{-0.5}${\bf \ times slower than a static }$%
\kappa -${\bf clock.}

Let $v$ ($\left| v\right| <1$) and $l$ be a real numbers, and $k_i$ be a
natural.

Let for the coordinate system $R^\mu $ of the frame reference $\left( \Re 
{\bf a}\widetilde{H}\right) $: at every moment $t$ the coordinates of:

${\bf b}$ be: $\left\langle x_{{\bf b},1}+v\cdot t,x_{{\bf b},2},x_{{\bf b}%
,3},\ldots ,x_{{\bf b},\mu }\right\rangle $,

${\bf g}_j$ be: $\left\langle x_{j,1}+v\cdot t,x_{j,2},x_{j,3},\ldots
,x_{j,\mu }\right\rangle $,

${\bf b}_j$ be: $\left\langle x_{j,1}+v\cdot t,x_{j,2}+l/\left( k_1\cdot
\ldots \cdot k_j\right) ,x_{0,3},\ldots ,x_{0,\mu }\right\rangle $,

for all $q_i$: if $q_i\in \Im $ then the coordinates of

$q_i$ be $x\left\langle _{i,1}+v\cdot t,x_{i,2},x_{i,3},\ldots ,x_{i,\mu
}\right\rangle $,

$\widetilde{T}$ be $\left\langle {\left\{ {\bf g}_1,A,{\bf b}_1\right\} ,\
\left\{ {\bf g}_2,A,{\bf b}_2\right\} ,...,\left\{ {\bf g}_j,A,{\bf b}%
_j\right\} ,\ ...\ }\right\rangle $.

In this case from Theorem 9:

$\Im $ is $ISS\left( {\bf b},\widetilde{T}\right) $.

Hence,{\bf \ the internal stableness survives for the uniform in-line motion.%
}

Let:

1) for the coordinate system $R^\mu $ of the frame reference $\left( \Re 
{\bf a}\widetilde{H}\right) $: at every moment $t$:

${\bf b}$ : $\left\langle x_{{\bf b},1}+v\cdot t,x_{{\bf b},2},x_{{\bf b}%
,3},\ldots ,x_{{\bf b},\mu }\right\rangle $,

${\bf g}_j$ : $\left\langle x_{j,1}+v\cdot t,x_{j,2},x_{j,3},\ldots
,x_{j,\mu }\right\rangle $,

${\bf b}_j$ : $\left\langle x_{j,1}+v\cdot t,x_{j,2}+l/\left( k_1\cdot
\ldots \cdot k_j\right) ,x_{0,3},\ldots ,x_{0,\mu }\right\rangle $,

for all $q_i$: if $q_i\in \Im $ then the coordinates of

$q_i$ : $\left\langle x_{i,1}+v\cdot t,x_{i,2},x_{i,3},\ldots ,x_{i,\mu
}\right\rangle $,

$\widetilde{T}$ be $\left\langle {\left\{ {\bf g}_1,A,{\bf b}_1\right\} ,\
\left\{ {\bf g}_2,A,{\bf b}_2\right\} ,...,\left\{ {\bf g}_j,A,{\bf b}%
_j\right\} ,\ ...\ }\right\rangle $,

$C$ : $\left\langle C_1,C_2,C_3,\ldots ,C_\mu \right\rangle $,

$D$ : $\left\langle D_1,D_2,D_3,\ldots ,D_\mu \right\rangle $,

$t_C=\left[ {\bf b}^{\bullet }C\mid \Re {\bf a}\widetilde{H}\right] $,

$t_D=\left[ {\bf b}^{\bullet }D\mid \Re {\bf a}\widetilde{H}\right] $;

2) for the coordinate system $R^{\mu \prime }$ of the frame reference $%
\left( \Im {\bf b}\widetilde{T}\right) $:

$C$ : $\left\langle C_1^{\prime },C_2^{\prime },C_3^{\prime },\ldots ,C_\mu
^{\prime }\right\rangle $,

$D$ : $\left\langle D_1^{\prime },D_2^{\prime },D_3^{\prime },\ldots ,D_\mu
^{\prime }\right\rangle $,

$t_C^{\prime }=\left[ {\bf b}^{\bullet }C\mid \Im {\bf b}\widetilde{T}%
\right] $,

$t_D^{\prime }=\left[ {\bf b}^{\bullet }D\mid \Im {\bf b}\widetilde{T}%
\right] $.

In this case from I, II, III:

\begin{center}
$\left( t_D^{\prime }-t_C^{\prime }\right) ^2-\left( D_1^{\prime
}-C_1^{\prime }\right) ^2-\left( D_2^{\prime }-C_2^{\prime
}\right)^2-\ldots-\left( D_\rho ^{\prime }-C_\mu ^{\prime }\right) ^2=$\\

=$\left( t_D-t_C\right) ^2-\left( D_1-C_1\right) ^2-\left(
D_2-C_2\right)^2-\ldots -\left( D_\rho -C_\mu \right) ^2$.\\
\end{center}

From above{\bf \ the Lorentz transformations} are obtained.

\section{RESUME}

The clock-like structure can be constructed from the recorders, and the
following results are obtained from I, II and III:

First, all such clocks have got the same direction, i.e. if the event,
expressed by the sentence $A$, precedes to the event, expressed by the
sentence $B$, with respect to any such clock, then it is the same for all
other such clocks.

Second, the Time, defined by such clocks, proves irreversible, i.e. no the
recorder can obtain the information, that a certain event has taken place,
before it has actually taken place. Thus, nobody can return back into the
Past Times or obtain the information from the Future Times.

Third, the set of recorders has been embedded in the metric space by some
natural method; i.e. all metric space axioms are obtained from I, II and III.

Fourth, if this metric space proves to be the Euclidean space, then the
corresponding recorders ''space-time'' obeys the Poincare complete group
transformations. I.e. in this case the Special Theory Relativity follows
from the logical properties of the information. If this metric space is not
Euclidean, then any non-linear geometry exists on the space of the
recorders, and any variant of the General Relativity Theory can be realized
on this space.

Therefore, the principal time properties - the one-dimensionality and the
irreversibility -, the space metric properties and the spatial-temporal
principles of the theory of the relativity are deduced from I, II, and III.
Hence, if you have got any set of the objects, which able to get, to keep or
to give any information, then ''the time'' and ''the space'' are inevitable
on this set. And it is all the same: or this set is in our world or this set
is in any other worlds, in which the spatial-temporal structure does not
exist initially. Hence the spatial-temporal structure arises from the logic
properties of the information.

{\bf The transformations of the complete Poincare group are obtained from
the logic properties of a information.}

\section{Acknowledgements}

Special thanks to Prof. Emilio Panarella.

\end{document}